\documentclass[doublecol]{epl2} 

\title{Bridges in three-dimensional granular packings: experiments and simulations}
\shorttitle{Bridges in $3d$ granular packings} 

\author{Y. X. Cao\inst{1} \and B. Chakrabortty\inst{2} \and G. C. Barker\inst{3}\thanks{E-mail: \email{gary.barker@ifr.ac.uk}} \and A. Mehta\inst{2}\thanks{E-mail: \email{anita@bose.res.in}} \and Y. J. Wang\inst{1}\thanks{E-mail: \email{yujiewang@sjtu.edu.cn}}}
\shortauthor{Y. X. Cao \etal}

\institute{                    
  \inst{1} Physics Department, Shanghai Jiao Tong University - 800 Dongchuan Rd. Shanghai 200240, China\\
  \inst{2} Theory Department, S N Bose National Centre - Block JD Sector III, Salt Lake, Calcutta 700098, India\\
  \inst{3} Institute of Food Research, Norwich Research Park - Norwich NR4 7UA, UK
}

\abstract{
In this Letter, we present the first experimental study of bridge structures in three-dimensional dry granular packings. When bridges are small, they are predominantly `linear', and have an exponential size distribution. Larger, predominantly `complex' bridges, are confirmed to follow a power-law size distribution. Our experiments, which use X-ray tomography, are in good agreement with the simulations presented here, for the distribution of sizes, end-to-end lengths, base extensions and orientations of predominantly linear bridges. Quantitative differences between the present experiment and earlier simulations suggest that packing fraction is an important determinant of bridge structure. }

\begin{document}

\maketitle


The study of random granular packings remains an active research field~\cite{torquato}. For packings containing frictional grains, it is now well established that cooperative structures such as bridges, are ubiquitous: these are defined as collective structures where neighboring grains rely on each other for mutual stability~\cite{het}. In other words, bridges are structures within a random close packed deposit that are inconsistent with the results of sequential deposition. The word ‘bridge’ is descriptive (arch would be equally valid) and is used widely within powder and bulk processing. Bridges are a measureable property of complex three dimensional granular structures and so their investigation can help build a more complete picture of the relationship between deposition processes and deposit structures.

\begin{figure}[h]
\onefigure[scale=0.32,trim=1cm 1cm 1cm 1cm]{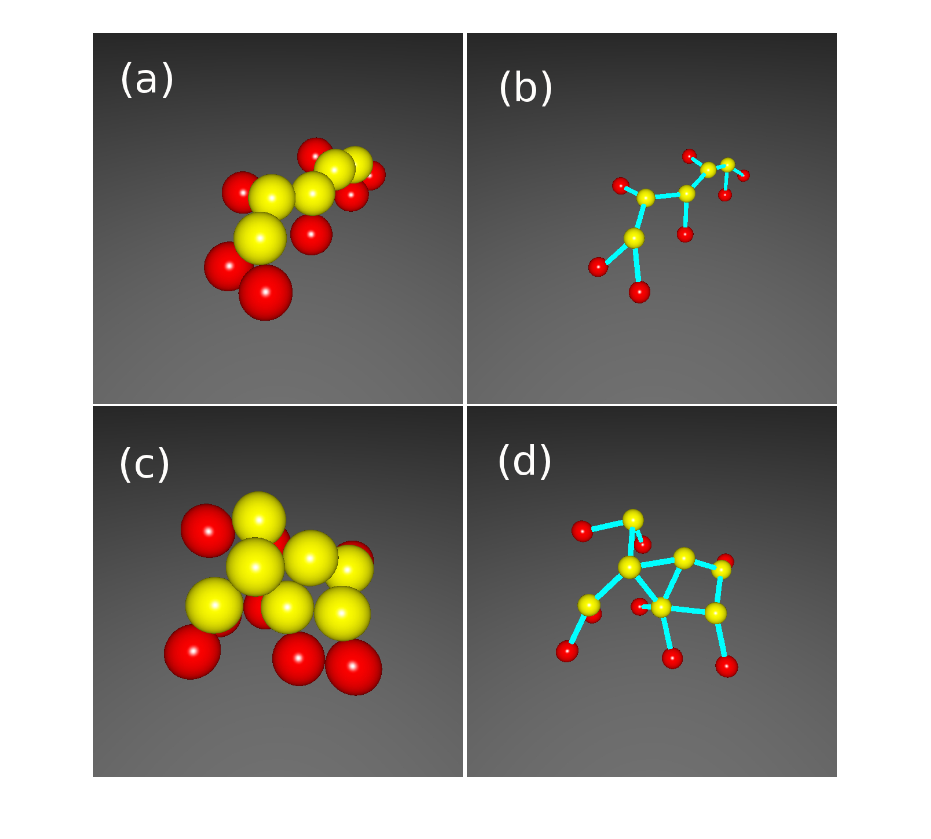}
\caption{(a) A typical linear bridge. (b) The backbone of the linear bridge shown in (a). (c) A typical complex bridge. (d) The backbone of the complex bridge shown in (c).}
\label{lincomp}
\end{figure}

Apart from their intrinsic interest, bridges are intimately related to the phenomenon of compaction \cite{ambook}; it has been shown that much of the compaction to high densities in shaken packings occurs via the cooperative dynamics of bridge collapse \cite{gcbampre}. Bridges can be further classified \cite{bridges} as linear or complex (see Fig.~\ref{lincomp}), depending on the topology of their backbones or contact networks: complex bridges have backbones with loops and/or branches, while linear bridges do not. Their dynamical implications are equally interesting: it has been suggested that grain motion close to the jamming transition \cite {liunagel} is via the motion of `dynamic linear chains' \cite{bebi, keys}, which are akin to linear bridges \cite{bridges}. Force networks in anisotropically sheared static packings \cite{howell, majmudar} show fractal dimensions similar to those of linear bridges formed under gravity, thus reinforcing connections that have already been made \cite{ambook, bridgeid, pug} between linear bridges and force chains. It should be emphasised that these connections are still somewhat qualitative, and that the precise characterisation of the relationship between force chains and bridges is an important and emergent research area. That there must be such a connection is evident: bridges are structures within a granular packing, whereas force chains involve combinations of  complex interparticle forces constrained by the structure. While the common element is the underlying contact network, this alone does not define force chains --- in that many force networks  are consistent with a mechanically stable structure containing bridges. From this point of view, studies of force chains and bridges are necessarily complementary in nature, and the study of bridges forms a valuable constraint on possible structures involving force chains.

Recent experimental advances have allowed for the non-invasive imaging of structure in dry granular media, using for example magnetic resonance imaging or positron emission tomography  \cite{MRI, PET}.  Although interesting attempts have been made to use these tools to characterise force chain distributions, they rely on indirect geometric measures of, say force chain lengths \cite{sanfratello}, rather than direct measurements of force. Bridge structures have also recently been probed experimentally in colloidal packings \cite{loadbear} and compared with the results of computer simulations of shaken (dry) granular media \cite{bridges}; while the experimental results are of interest in and of themselves, it has to be remembered that colloids are governed by thermal energy, while temperature does \textit{not} govern the dynamics of dry granular media. Instead, a perturbation such as shaking or tapping has to be applied \cite{rmp} in order to generate particulate motion in dry granular packings; such athermal perturbation could conceivably generate its own particularities in bridge structure, absent in the colloidal case. It is therefore important to characterise the statistics of bridges in the steady state of shaken \textit{dry} granular packings in $3d$ and to compare them with the results of computer simulations modelling \textit{exactly the same} physical situation; this is the main purposes of this Letter, where we use  direct and non-invasive tomographic measurements to characterise bridge structures
in dry granular packings.

\begin{figure}[h]
\onefigure[scale=0.27,trim=2cm 1cm 2cm 1cm]{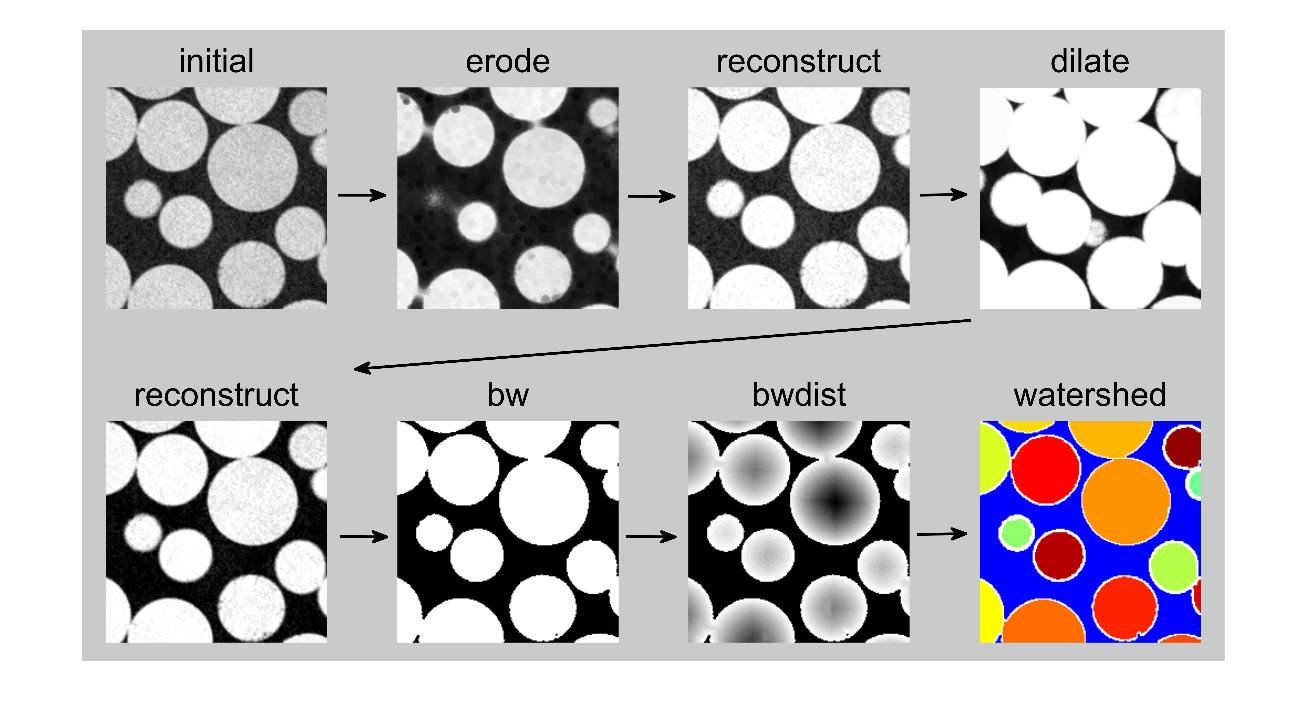}
\caption{(color online) Image processing steps from the raw reconstructed image to the segmented image using a marker-based watershed algorithm.}
\label{waterex}
\end{figure}

In the experiments, two packings (one monodisperse (Duke Scientific, USA) and one polydisperse) of glass beads  were used, with packing fractions of $0.623$ and $0.597$ respectively. The glass beads had diameters of $200\pm 15$ $\mu$m and $300\pm 50$ $\mu$m respectively: there were  $\sim 18000$ beads in the monodisperse packing and  $\sim 5000$ beads in the polydisperse packing. Both packings were tapped for more than $10,000$ cycles using a commercial shaker to ensure that a steady state had been reached. The tapping protocol involved a single 30-Hz sine wave at a rate of 1 Hz with an effective tapping amplitude of $2.82$ g. An X-ray microtomography machine (MicroXCT-200, Xradia Inc., USA) was used to make structural measurements, with $1200$ projection images taken on the samples. The effective spatial resolution of the detector was $6.88 \times 6.88$ $\mu m^2$ after optical magnification ($2 \times$). The tomography-reconstructed $3d$ images were analyzed by a marker-based watershed imaging segmentation technique~\cite{Fu, Tsukuhara} to ensure an accurate determination of contacting neighbours which is crucial for the identification of bridges. The watershed algorithm represents a grey-scale image as a topological surface, where the grey-scale value of each pixel is interpreted as its `altitude'. The initial image is separated into a binary image which comprises a `solid' area and a background; the key step is to transform each solid area into a single `catchment basin'. Before this process is carried out, the standard `erosion' and `dilation' steps were carried out to remove the noise. Afterwards, a `reconstruction' step is invoked before the distance transform (`bwdist') is performed to compute the nearest distance between two pixels corresponding to the background, thus identifying the limits of the solid phase, or `catchment basin'. The edges of each solid phase, i.e. each grain, are identified as a watershed ridge line (see white lines in last colour panel of Fig.~\ref{waterex}).

The identification of bridges in experiment and simulation followed the overall algorithm used in \cite{bridges, bridgeid, luis, matthew}. Sphere coordinates at the end of the stabilization phase are transformed into a list of at least three contacts for each particle. Each particle in a contact list is next identified with a unique set of three other particles that provide its supporting base, using a criterion favouring the stabilizing triplet with the lowest centroid. Once these stabilizing triplets are identified, one looks for sets of particles that appear in each other's stabilizing triplets: these are clearly particles which are mutually stabilizing, and thus identified with bridge particles. Finally, clusters of mutual stabilizations can be identified, using a linked list structure as in other aggregation applications, to reveal a unique set of bridges in a static close packing. The statistics of these clusters can be used to get the kind of information on bridge structure such as sizes, orientations and topologies, referred to below.

\begin{figure}[h]
\onefigure[scale=0.32,trim=0cm 0cm 0cm 0cm]{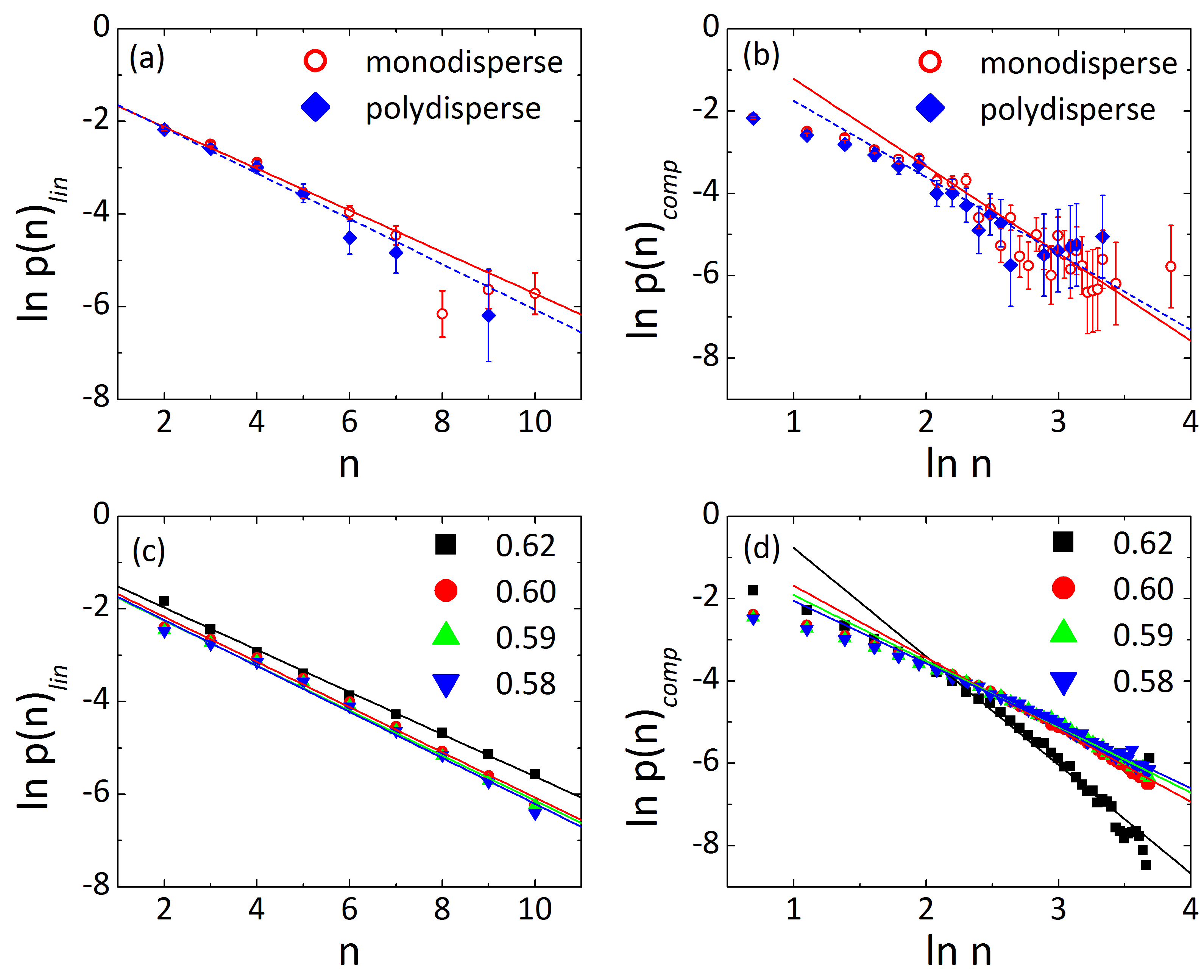}
\caption{(color online) (a) Log-linear plot of experimental size distributions $P(n)_{lin}$ of linear bridges: the full and dashed lines represent  exponential-law fits for mono- and polydisperse beads respectively. The corresponding exponent values are   $\alpha = 0.45 \pm 0.02$ (monodisperse) and $\alpha = 0.49 \pm 0.04$ (polydisperse). (b) Log-log plot of experimental size distributions $P(n)_{comp}$ for all bridges. The power-law fits to the largely complex bridges for $n \geq 7$ yield $\tau = 2.1 \pm 0.2$ (monodisperse) and $\tau = 1.9 \pm 0.4$ (polydisperse). (c) Log-linear plot of size distributions $P(n)_{lin}$ for linear bridges in packings at different densities $\phi$ from computer simulations of shaken monodisperse spheres. The black line is an exponential-law fit as in panel (a) to the results for $\phi=0.62$, yielding $\alpha = 0.46 \pm 0.01$. For the three lower densities, data collapse is obtained, giving $\alpha = 0.49 \pm 0.02$. (d) Log-log plot of  size dis
 tributions $P(n)_{comp}$ for all bridges in packings at different densities $\phi$,  from computer simulations of shaken monodisperse spheres. The black line represents a power-law fit for $n \geq 7$ as in panel (b) for $\phi=0.62$, giving  $\tau = 2.6 \pm 0.1$. For $\phi=0.60,0.59,0.58$, $\tau = 1.8 \pm 0.1, 1.7 \pm 0.1, 1.5 \pm 0.1$ respectively. The exponent $\tau$ appears to vary strongly with packing fraction.}
\label{linearex}
\end{figure}

Our experimental findings suggest that linear bridges predominate for sizes of up to $n \approx 10$, which are characterised by a simple exponential distribution $P(n)_{lin} \sim e^{-\alpha n}$ (Fig.~\ref{linearex}(a)). For larger sizes $n$, complex bridges predominate, which are in turn characterised by a power-law distribution $P(n)_{comp} \sim n^{-\tau}$ (Fig.~\ref{linearex}(b)). These results are robust to the presence of polydispersity: remarkably, even the associated exponents agree --- within error bars--- in the two cases, with $\alpha =0.45 \pm 0.02$ and $\tau = 2.1 \pm 0.2$ (monodisperse) and  $\alpha =0.49 \pm 0.04$ and $\tau =1.9 \pm 0.4$ (polydisperse).

The above results are in broad qualitative agreement with earlier simulation results on simple and complex bridges~\cite{bridges}, which is in itself quite good. There are, however, quantitative differences in the values of the exponents measured. We speculate that this could be due to the difference in packing fractions $\phi$ considered in the earlier simulations and in the current experiment ($0.56$ and $0.62$ respectively); since packing fraction is a fundamental structural descriptor of granular media, it could reasonably lead to quantitative, if not qualitative differences in the size distribution of granular bridges. The simulations described below investigate this issue, and confirm this dependence.

Accordingly, we generated configurations corresponding to $\phi = 0.58$, $0.59$, $0.60$ and $0.62$ using a well-established hybrid Monte Carlo sphere-shaking algorithm~\cite{amgcbprl}. The  simulations were performed on $1630$ spheres in a rectangular cell with lateral periodic boundaries and a hard disordered base, using $100$ different random initial configurations per shaking amplitude. Size checks were performed, and qualitatively similar configurations were obtained for a couple of different system sizes. Care was taken to use stable configurations in the steady state, saving about $200$ stable configurations (picked out every $500$ cycles to avoid correlation effects) for bridge identification and analysis. Our findings are shown in  Figs.~\ref{linearex}(c) and (d) where we find that there continues to be excellent qualitative agreement between experiment and simulation. Quantitatively, we observe an interesting trend: the agreement between experiment and simulation gets better as the corresponding packing fractions converge. In fact for $\phi = 0.62$, our simulations yield values of $\alpha = 0.46 \pm 0.01$ and $\tau = 2.6 \pm 0.1$ (see black line in Figs.~\ref{linearex}(c) and (d)). The  perfect agreement within error bars for the $\alpha$ values between experiments and simulations conducted at the \textit{same} packing fraction suggests that this is the most important parameter controlling the behaviour of linear bridges. We also notice that for the three lower packing fractions, the exponent $\tau$ show a consistently increasing trend as a function of $\phi$. However, at $\phi = 0.62$, the exponent $\tau$ from experiment is more compatible with the trend than that from simulation. This is possibly due to the onset of ordering in the simulations \cite{torquato}.

\begin{figure}[h]
\onefigure[scale=0.32,trim=1cm 1cm 1cm 1cm]{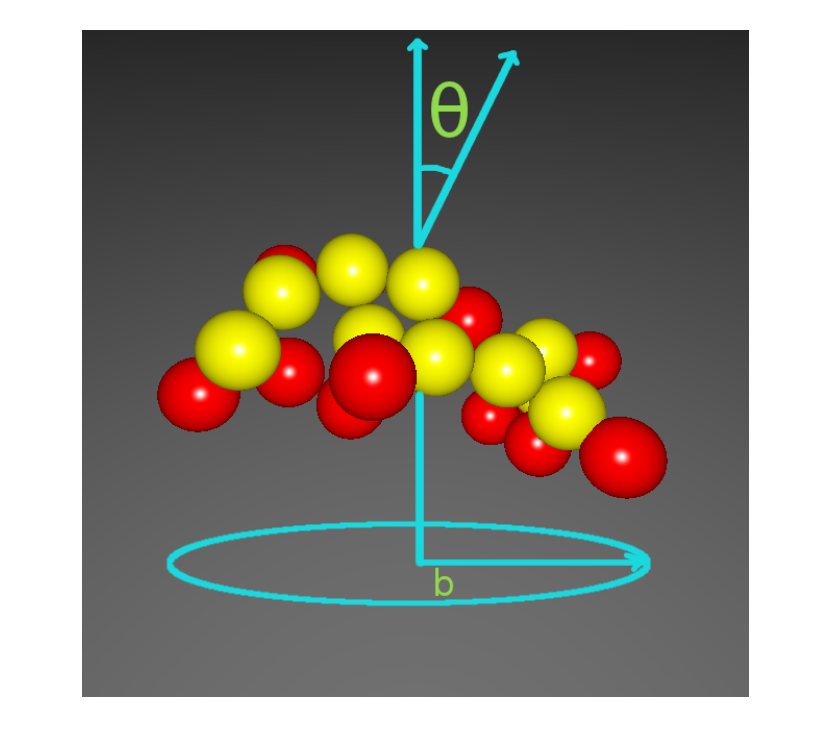}
\caption{(color online) The base extension of a bridge.}
\label{basex}
\end{figure}

All these trends persist in measurements of the other bridge structure descriptors considered below; in the following we focus on linear bridges, leaving the detailed examination of complex bridges to future work. The first of our descriptors is the base extension, whose definition we review. If all possible connected triplets of base particles for a particular bridge are considered,  the vector sum of their normals is defined to be the direction of the {\it main axis} of the bridge, typically inclined at some angle $\Theta$ (the orientation angle) to the $z$-axis. The base extension $b$ (see Fig.~\ref{basex}) is defined as the radius of gyration of the base-particles {\it about the $z$-axis}, and is a measure of the spanning or jamming potential of a bridge \cite{bridges}.

\begin{figure}[h]
\onefigure[scale=0.32,trim=0cm 0cm 0cm 0cm]{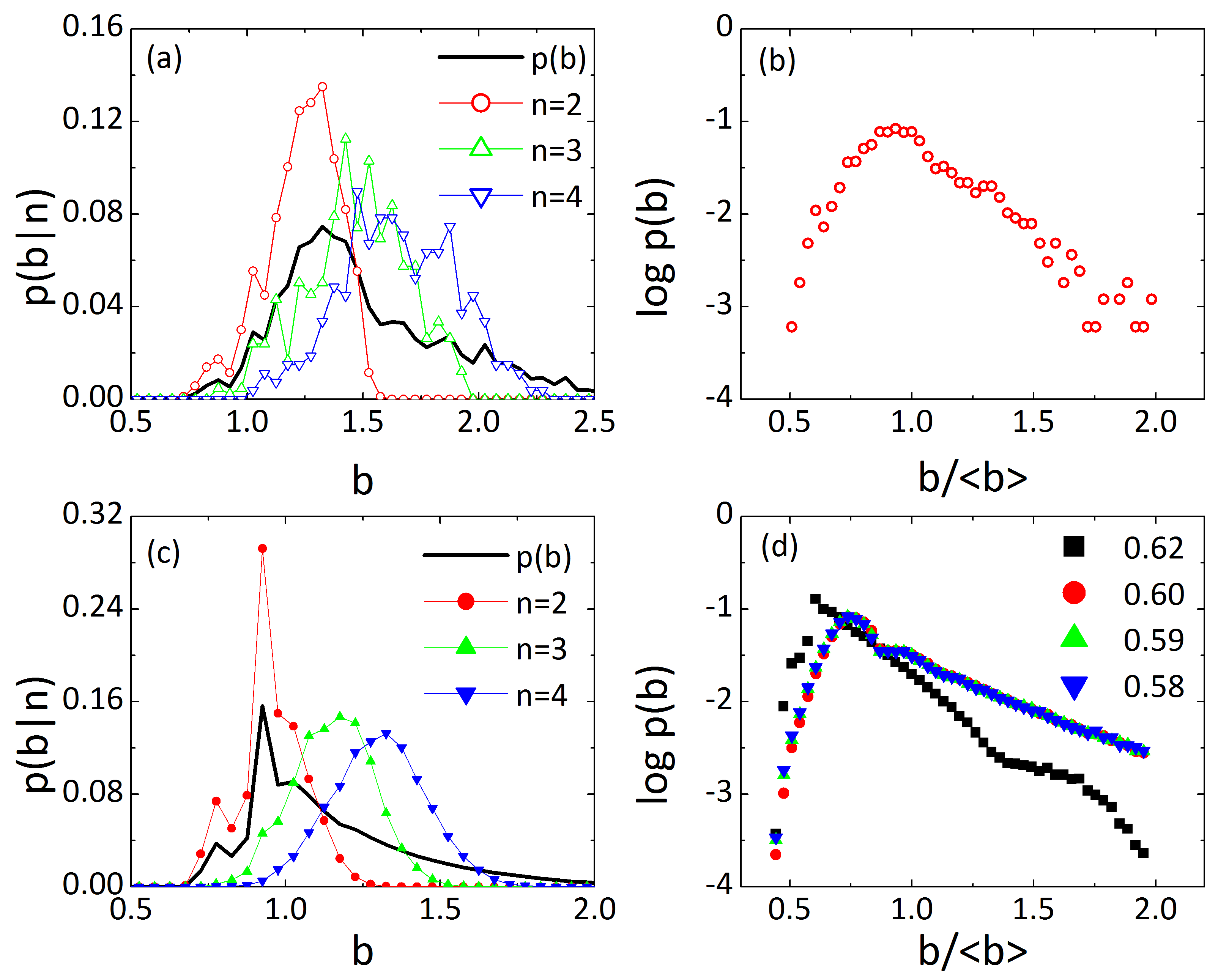}
\caption{(color online) Probability distributions, for monodisperse beads, of base extensions $b$ for linear bridges. (a) Experimental plot of $p(b|n)$ vs. $b$ for different $n$. (b) Experimental plot of the log of the cumulative distribution, log $(p(b))$ vs. $b/<b>$. (c) Plot of $p(b|n)$ vs. $b$ for different $n$, generated from computer simulations of shaken monodisperse spheres. (d) Plot of log $(p(b))$ vs. $b/<b>$ from computer simulations of monodisperse sphere packings at different densities.}

\label{pbn}
\end{figure}

In Fig.~\ref{pbn}, we present experimental and simulation results obtained for the base extension distributions of linear bridges. Figs.~\ref{pbn} (a) and  (c) correspond respectively to experimental and simulation plots of $p(b|n)$ (normalised probability distributions of $b$ conditional on bridge size $n$) vs. $b$ for different bridge sizes $n$. They are qualitatively similar, showing sharply peaked distributions which flatten out with increasing bridge sizes, as seen in earlier simulations \cite{bridges}. Figs.~\ref{pbn} (b) and (d)  correspond respectively to experimental and simulation plots of log $p(b)$ (cumulative probability distribution of $b$) vs. the normalised variable $b/<b>$, with $<b>$ the mean extension of bridge bases. Both show the exponential tail in the distribution function noted in the results of earlier simulations \cite{bridges}, suggesting that bridges with small base extensions are not favored. This result appears to be robust both with respect to polydispersity in the experimental results and packing fractions in the simulations. It also reinforces earlier suggestions \cite{ambook} of deep connections between force chains and linear bridges: the cumulative distributions of force chains in anisotropically sheared granular systems~\cite{majmudar, mueth} as well as MD simulations of analogous particle packings~\cite{sno, ohern} show very similar exponential tails. Recent simulations have directly confirmed the connection between force chains and linear bridges, by suggesting that forces are principally transmitted by particles in bridges~\cite{pug}.

\begin{figure}[h]
\onefigure[scale=0.32,trim=0cm 0cm 0cm 0cm]{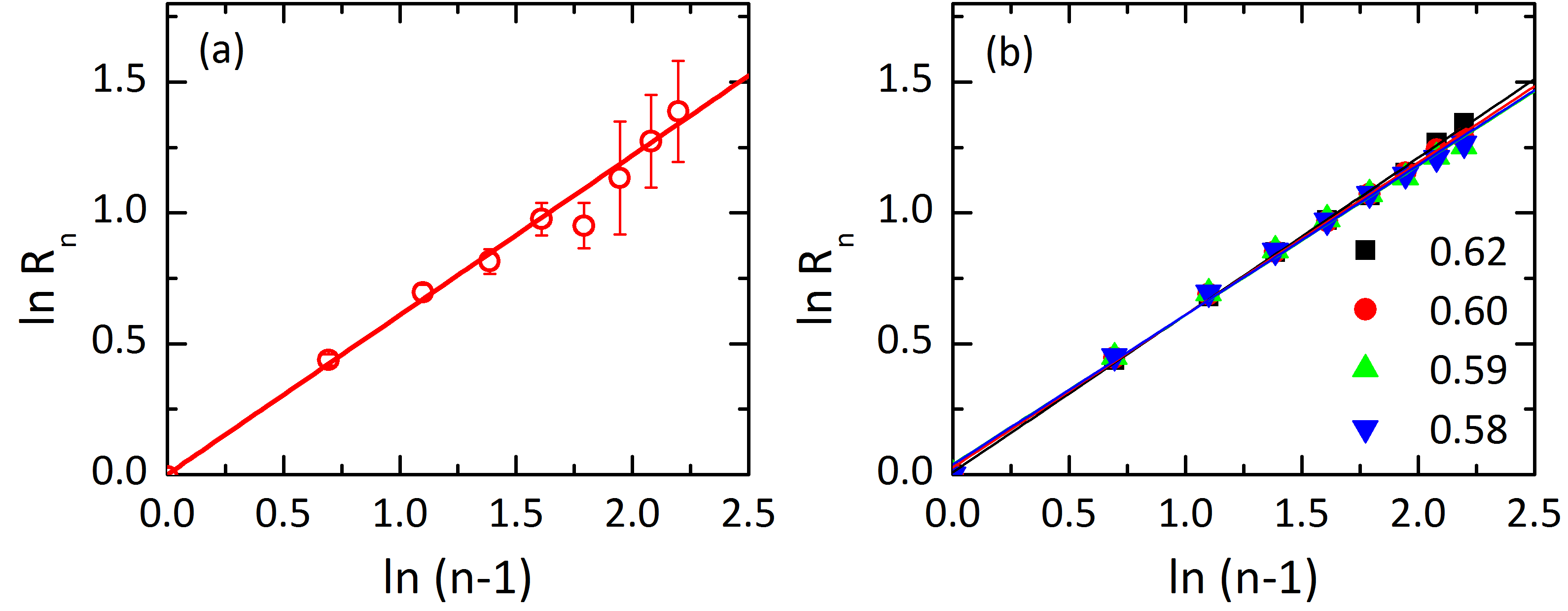}
\caption{(color online) Plot of the (log of the) end-to-end length, $\ln R_{n}$ vs. $\ln (n-1)$ for linear bridges in monodisperse sphere packings. The full lines represent power-law fits.  For the experimental plot (a) $\nu = 0.61 \pm 0.02$, while (b) simulations for all the densities give $\nu = 0.60 \pm 0.01$.}
\label{Rn}
\end{figure}

Another important quantity related to a linear bridge of size $n$ is obviously its `span', i.e. its rms end-to-end length $R_{n}$. Our results for this are presented in  Fig.~\ref{Rn}(a) (experiment) and Fig.~\ref{Rn}(b)(simulations). As expected, we find the scaling law $R_{n} \sim n^{\nu}$ in both experiment and simulation.  We obtain exponent values $\nu = 0.61 \pm 0.02$ from experiment and $\nu = 0.60 \pm 0.01$ from simulations. Again, the agreement between experiment and simulations is remarkable; additionally, we do not observe a strong dependence on packing fraction. Given that $\nu \sim 0.59$ for a  $3d$ self-avoiding random walk, this suggests that the linear bridges that we have examined here look -- within error bars -- \textit{exactly} like self-avoiding walks in three dimensions. 

\begin{figure}[h]
\onefigure[scale=0.32,trim=0cm 0cm 0cm 0cm]{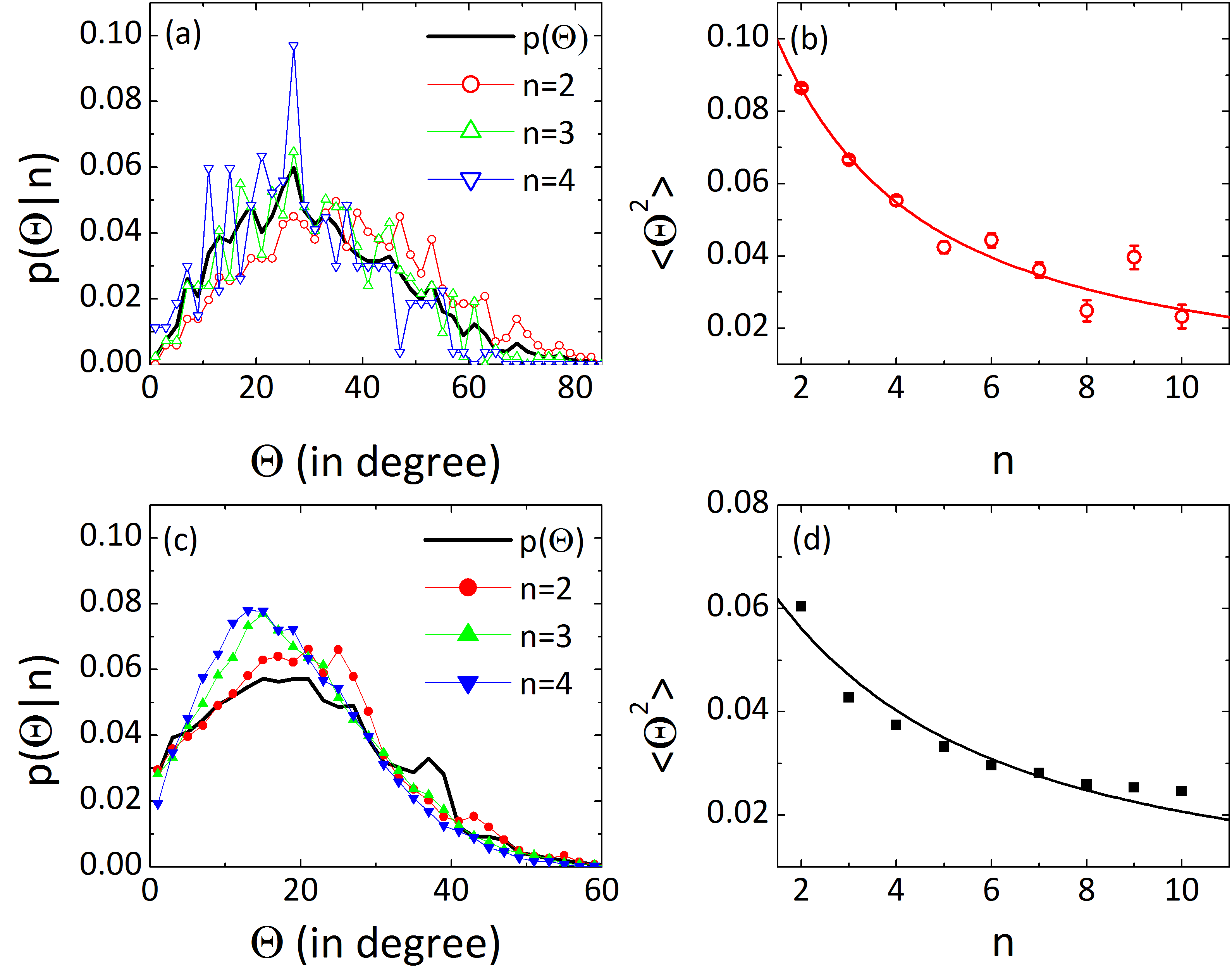}
\caption{(color online) Orientational distributions of linear bridges in monodisperse systems from experiment and simulation. (a) Plot of the orientational distribution conditional on size $n$, $p(\Theta|n)$  vs. $\Theta$, obtained experimentally. (b) Experimental plot of the variance of $\Theta$, $\left\langle\Theta^{2}\right\rangle$ as a function of $n$: the full line shows the fit to  Eq. \ref{eq1}. (c) Plot of the orientational distribution conditional on size $n$, $p(\Theta|n)$ vs. $\Theta$, obtained from computer simulations of shaken spheres.  (d) Computer simulation generated plot of the variance of $\Theta$, $\left\langle\Theta^{2}\right\rangle$ as a function of $n$: the full line shows the fit to  Eq. \ref{eq1}, as in (b).}

\label{ptheta}
\end{figure}

Finally, we examine via experiment and simulation, the normalised distribution for the mean angle $\Theta$ made by a linear bridge with the $z$-axis, as well as that of its variance \cite{bridges} in Fig.~\ref{ptheta}. In Figs.~\ref{ptheta}(a) and (c), we plot respectively experimental and numerical results for $p(\Theta|n)$, the orientational distribution conditional on $n$.  The cumulative distributions $p(\Theta)$, also plotted in these figures, closely resemble each other as well as confirming earlier results \cite{bridges}. Notable features are a peak around $20^{o}$ as well as the decrease of  $\Theta$  with increasing bridge size, which \cite{bridges} suggests that larger linear bridges form domes.  In Figs.~\ref{ptheta}(b) and (d) we plot the variance of the mean orientational angle, $\left<\Theta^{2}\right>$, against size $n$. According to~\cite{bridges},  $\left<\Theta^{2}\right>$ obeys

{\small
\begin{equation}
\left\langle\Theta^{2}\right\rangle(s)=2\sigma_{eq}^{2}\frac{as-1+e^{-as}}{a^{2}s^{2}}+\left(\sigma_{0}^{2}-\sigma_{eq}^{2}\right)\frac{(1-e^{-as})^{2}}{a^{2}s^{2}},
\label{eq1}
\end{equation}
}
where $\sigma_{eq}^{2}$ is  the equilibrium value of the variance of the link angle. This is plotted as the full line in the figures, where the symbols represent results for experiment and simulation respectively. These look similar to each other, and also to the results of earlier simulations \cite{bridges}.

Before concluding, we mention that while we hope our investigations	 of bridge structures might shed some light on possible relationships between structural signatures and force networks in the future, we do not for the present include any explicit force information directly.  Our investigations of structure are \textit{in addition} to ongoing force chain investigations \cite{karen}, since stable structures are a constraint on force chains, and the two taken together might lead some day to the emergence of a holistic picture of heterogeneities in granular systems.

The experiments reported in this Letter show satisfying qualitative agreement with the results of present and earlier \cite{bridges} simulations, in the sense that the forms of distributions of quantities ranging from sizes to orientations and base extensions are robustly
the same in every case. This is already quite good for such a complex field, and suggests that there is truth in the idea of bridges being
good characterisers of spatial heterogeneities in granular media. Quantitative differences, where these exist, between the current experiment and earlier simulations \cite{bridges} have been successfully ascribed to the fact that the data were taken at rather different packing fractions $\phi$: the simulations reported in this paper have been conducted at a range of packing fractions and manifest this dependence, such that there is excellent agreement between simulation and experiment at matching $\phi$. This reinforces the intuitive idea that packing fractions should influence the details of bridge structure (while leaving the qualitative features unchanged); higher packing fractions, for instance, probably constrain linear bridges to be fully three-dimensional rather than leaving them the choice to be planar. We might expect that for complex bridges, the factors at play (apart from the robust power-law size distribution) may indeed be more complex: since these are branched structures, we might expect that the coordination number, might have an important role to play in precise quantitative measurements of structure, as indicated by our measurements of the exponent $\tau$. We hope to carry out a detailed study of complex bridges in future work.

\acknowledgments
Some of the initial work has been carried out at the BL13W1 beamline of Shanghai Synchrotron Radiation Facility (SSRF), and the work is supported by the Chinese National Science Foundation No. 11175121, Shanghai Pujiang Program (10PJ1405600), program for New Century Excellent Talents (NCET) in University, National Basic Research Program of China (973 Program; 2010CB834300).

\end{document}